\documentclass[]{emulateapj}
\usepackage{natbib}
\usepackage{graphicx}
\usepackage{amsmath}
\usepackage{amssymb}

\begin{document}
\shorttitle{Axis Ratio Distribution of Early-type galaxies}
\shortauthors{Taysun Kimm \& Sukyoung K. Yi}

\title{Intrinsic Axis Ratio Distribution of Early-type galaxies from Sloan Digital Sky Survey}

\author{Taysun Kimm and Sukyoung K. Yi}
\affil{Department of Astronomy, Yonsei University, Seoul 120-749, Korea}
\email{E-mail:yi@yonsei.ac.kr}

\slugcomment{Resubmitted to ApJS: \today}

\begin{abstract}
Using Sloan Digital Sky Survey Data Release 5, we have investigated the
intrinsic axis ratio distribution (ARD) for early-type galaxies.
We have constructed a volume-limited sample of 3,922 visually-inspected
early-type galaxies
at $0.05 \leq z \leq 0.06$ carefully considering sampling biases caused
by the galaxy isophotal size and luminosity.
We attempt to de-project the observed ARD into three-dimensional types
(oblate, prolate, and triaxial), which are classified in terms of triaxiality.
We confirm that no linear combination of $randomly$-distributed axis ratios
of the three types can reproduce the observed ARD.
However, using Gaussian intrinsic distributions,
we have found reasonable fits to the data with preferred mean axis ratios for
oblate, prolate, and triaxial (triaxials in two axis ratios),
$\mu_o=0.44, \mu_p=0.72, \mu_{t,\beta}=0.92, \mu_{t,\gamma}=0.78$  
where the fractions of oblate, prolate and triaxial types are 
$\textrm{O:P:T}=0.29^{\pm0.09}:0.26^{\pm0.11}:0.45^{\pm0.13}$. 
We have also found that the luminous sample ($-23.3 < M_r \leq -21.2$)
tends to have more triaxials than the less luminous ($-21.2 < M_r <-19.3$)
sample does.
Oblate is relatively more abundant among the less luminous galaxies.
Interestingly, the preferences of axis ratios for triaxial types in the two
luminosity classes are remarkably similar.
We have not found any significant influence of the local galaxy number density
on ARD. We show that the results can be seriously affected by
the details in the data selection and type classification scheme. 
Caveats and implications on galaxy formation are discussed.
\end{abstract}

\keywords{
surveys --- method: statistical --- galaxies: elliptical and lenticular, cD --- galaxies: formation --- galaxies: kinematics and dynamics
}

\section{INTRODUCTION}

Ever since Hubble (1926) investigated the apparent flattenings of early-type galaxies,
numerous studies have attempted to pin down their intrinsic shapes.
Such efforts mainly focused on the apparent axis ratio not only because
it is easy to measure but also its distribution can be used to extract
the kinematics and may
even constrain the general formation history of galaxies.

Early studies using apparent axis ratio distribution (ARD)
were made on the assumption that early-type galaxies are composed of one type.
Assuming oblateness alone, Sandage, Freeman, \& Stokes (1970) suggested
the possibility of Gaussian or skewed binomial distribution of intrinsic ARD
using Reference Catalog of Bright Galaxies (RC1, de Vaucouleurs \& de Vaucouleurs 1964).
Binney (1978)'s approach assuming prolateness also successfully reproduced
the apparent ARD from RC1. The existence of triaxial galaxies
(Bertola \& Capaccioli 1975; Illingworth 1977) opened up a better chance to
reproduce the observed ARD.
Binggeli (1980) and Benacchio \& Bertola (1980) showed that the
apparent ARDs for 160 ellipticals from Revised-Shapley-Ames catalog of
bright galaxies (Sandage \& Tammann 1979) and 348 ellipticals from Strom \&
Strom (1978a,b,c) were well represented as a group of triaxial galaxies
assuming a fixed case of $\beta=(1+\gamma)/2$ where $\beta$ is the
ratio between the second longest to the longest axis and $\gamma$ is
the shortest to the longest axis.
In particular, Binney \& de Vaucouleurs (1981) aimed to reconstruct
the apparent ARD for RC2 (de Vaucouleurs et al. 1976) using Lucy's inversion
technique (Lucy 1974, see also Noerdinger 1979).
They considered oblate, prolate and triaxial (O,P,T) separately and still
found acceptable fits to the observed ARD.
The limitation of this technique, however,
is that it is difficult to constrain the ratios between O, P, T galaxies.

More recently, Fasano \& Vio (1991) concluded that a purely-biaxial model
cannot reproduce the small number of apparently-round galaxies.
This paucity however looks significantly different when samples are drawn
from different catalogs (e.g., RC1, RC2, or Revised-Shapley-Ames
Catalog of Galaxies).
This obviously results in disparate intrinsic distributions.
Furthermore, Lambas et al. (1992) found a reasonable fit to the sample
of 2,135 galaxies from the APM Bright Galaxy Survey (Maddox 1990)
assuming all ellipticals are triaxial
whose intrinsic distributions are two dimensional Gaussian.
The similar work done by Ryden (1992) presents consistent results.
These results support the assertion that the paucity of round galaxies
can be reproduced by a dominantly-triaxial galaxy population.
Considering its usefulness, it is important to accurately sample
the observed ARD.

Fortunately, Sloan Digital Sky Survey Data (Adelman-McCarthy et al. 2005)
allows us to study the apparent flattenings for a large number of galaxies.
Our effort to make a complete volume-limited sample
is one of the main factors distinguishing our work from previous studies.

We also note the interesting result of Davies et al. (1983)
that bright ellipticals may be more slowly rotating than faint ones.
Tremblay \& Merritt (1996) found that the apparent ARD is markedly
different for two luminosity classes ; the ellipticals brighter than
$M_B \simeq -20 $ are rounder than less luminous ones.
In particular, recent observations show that the division clearly
occurs at $M_B=-20.5$ (Rest \& van den Bosch 2001) and also support
previous works suggesting that bright galaxies show a core profile and
boxy isophotes while faint galaxies have a power-law profile and disky
isophotes (Bender 1988, Kormendy \& Bender 1996, Faber et al 1997).
Therefore we investigate on the effect of the galaxy luminosity on the ARD.

Another factor we should note for intrinsic shape of early-type
galaxies is an environmental dependence.
Dressler (1980) pointed out a density-morphology relation
which reflects the importance on the formation process.
We thus search for the connection between intrinsic shape and environment
for early types.

In this paper, we propose two simplifying assumptions;
(i) early-type galaxies are geometrically perfect ellipsoid,
(ii) they are randomly oriented.
We assume that early types consist of oblate, prolate and triaxial.
We believe this assumption makes our approach more realistic
than the previous models composed of only one or limited types.
On this basis we investigate the projection effect on the apparent ARD.
We simply describe the model distribution for oblate, prolate and triaxial in \S 2.
We introduce our sample selection with completeness tests in \S 3.
In \S 4, we investigate the intrinsic ARD for volume-limited samples.
In \S 5,  we analyze the intrinsic shape of two different luminosity samples.
Dependence of environment on ARD is investigated in \S 6.
In \S 7, we discuss the limitations of our approach.
Finally, we discuss results and their implication in \S 8.

\section{Analytic Apparent Axis Ratio Distribution}

We use an analytical description to calculate the probability distribution
(Franx et al. 1991; Binney \& Merrifield, 1998).
The probability of finding the apparent ellipticity ($\epsilon$) in the
interval ($\epsilon$, $\epsilon+d\epsilon$) is
\begin{equation}
p(\epsilon) d\epsilon = \left \{ \begin{array}{ll}
\frac{(1-e)\sqrt{e}} {\pi} \int_{\mu_2}^{\mu_1} \frac{\mu^2}{\sqrt{-h(\mu) h(e\mu)}}
   d\mu d\epsilon & \mathrm{for}~0\leq \epsilon \leq \epsilon_1,\\
0 & \mathrm{for}~\epsilon_1 \leq \epsilon \leq 1,
\end{array} \right.
\end{equation}
where $e=(1-\epsilon)^2$, $h(\tau)=(\tau - a^2)(\tau-b^2)(\tau -c^2)$ and
$\epsilon_1 = 1-b/a$.
$a$, $b$ and $c$ indicate three axes of an ellipsoid, respectively.
For galaxies with $b/a > c/b$ (oblate-triaxial),
$\mu_1$ and $\mu_2$, which depend on the galaxy shape, are
\begin{equation}
(\mu_1, \mu_2) = \left \{ \begin{array}{lll}
(b^2, b^2/e) & \mathrm{for} & 0 \le \epsilon \le \epsilon_2,\\
(b^2, a^2) & \mathrm{for} & \epsilon_2 < \epsilon \le \epsilon_3,\\
(c^2/e, a^2) & \mathrm{for} & \epsilon_3 < \epsilon \le \epsilon_1
\end{array}\right.
\end{equation}
where $\epsilon_3 = 1 - c/b$. For galaxies with $b/a \le c/b$ (prolate-triaxial),
$\mu_1$ and $\mu_2$ can be written as
\begin{equation}
(\mu_1, \mu_2) = \left \{ \begin{array}{lll}
(b^2, b^2/e) & \mathrm{for} & 0 \le \epsilon \le \epsilon_3,\\
(c^2/e, b^2/e) & \mathrm{for} & \epsilon_3 < \epsilon \le \epsilon_2,\\
(c^2/e, a^2) & \mathrm{for} & \epsilon_2 < \epsilon \le \epsilon_1
\end{array} \right.
\end{equation}
In order to illustrate the type (OPT)-dependence of the apparent ARD,
we adopt the classification scheme of Franx et al. (1991).
To classify the early-type systems, we use triaxiality (T),
\begin{equation}
T = \frac{1 - {\beta}^2}{1 - {\gamma}^2}
\end{equation}
and each type can be expressed as
\begin{displaymath}
\left\{\begin{array}{lll}
 \mathrm{oblate} &:& 0 \leq T < 0.25  \\
 \mathrm{triaxial} &:& 0.25\leq T < 0.75\\
 \mathrm{prolate} &:&0.75 \leq T \leq 1.0
\end{array} \right.
\end{displaymath}

\begin{figure}
\begin{center}
\includegraphics[width=8cm]{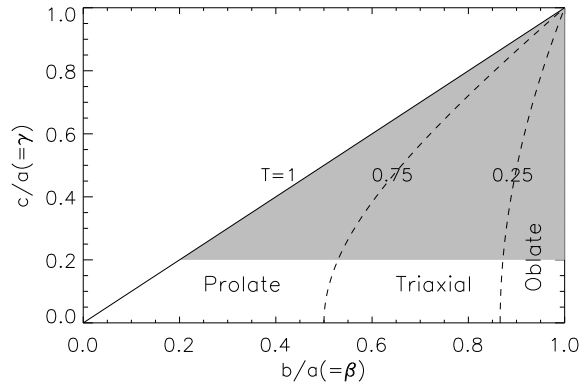}
\caption{Scheme of galaxy classification. In this study, we adopt Franx(1991)'s scheme.
Shaded area : Based on the observational constraints, we use axis ratios greater than 0.2.} \label{fig1}
\end{center}
\end{figure}

\begin{figure}
\begin{center}
\includegraphics[width=8cm]{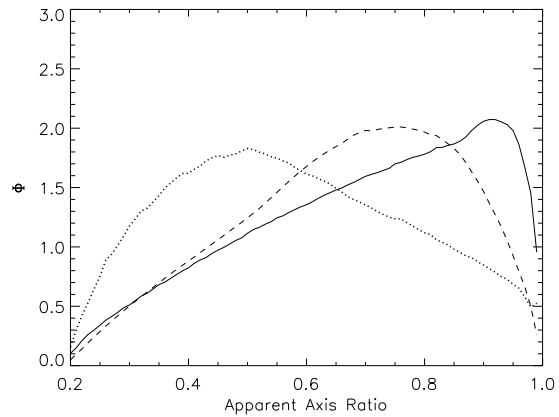}
\caption{
Apparent axis ratio distribution, $\Phi$ (i.e., the probability density
of the projected shapes of a group of galaxies)
for each type assuming a uniform intrinsic ARD.
Solid, dotted, and dashed lines correspond to oblate, prolate, and 
triaxial, respectively.} \label{fig2}
\end{center}
\end{figure}

In this study, we assume that there is no early-type galaxy with axis ratios
smaller than 0.2 because such systems are rarely observed.
In Fig. \ref{fig1} we display the classification scheme.
Fig. \ref{fig2} shows the simplest special case of the apparent ARD,
that is based on ({\it uniformly-distributed} intrinsic axis
ratios. We mean {\it no preferred values of the intrinsic ratios} by ``uniformly-
distributed''.
The numbers of the samples of three (OPT) types simulated are
normalised to be the same, hence unbiased by the area difference between the types
in Fig.\ref{fig1}. If an observed ARD has a large number of round galaxies near 1, we can
deduce that oblate is the main component, for example.

\section{SDSS SAMPLE SELECTION AND DATA ANALYSIS}

The SDSS provides a large homogeneous database. There are approximately
28,000 galaxies within $0.05 \leq z \leq 0.06$,  an excellent sample
for studying ARD statistically.
In this section, we describe our data selection scheme and completeness tests.

\subsection{Morphological Classification}

A well-defined criterion for morphological classification is necessary
to study the ARD of early-type galaxies.
Because visual inspection of all galaxies is
extremely time-consuming and still subjective,
we adopted the SDSS pipeline parameter
$fracDev$, which indicates the fraction of the brightness profile that
can be explained by the de Vaucouleurs profile (de Vaucouleurs 1948)
as follows:
\begin{equation}
f_{composite} = fracDev f_{deV} + (1-fracDev) f_{exp}
\end{equation}
where $f_{deV}$ indicates de Vaucouleurs fluxes. We assume that
a conservative sample of early-type galaxies have $fracDev\geq0.95$
in all 3 bands, $g'r'i'$, following the practice of Yi et al. (2005).
Using this $fracDev$ parameter,
we compile 4,994 galaxies within $0.05 \leq z \leq 0.06$.

\subsection{Data analysis}

\subsubsection{Luminosity Dependence}
\begin{figure}
\begin{center}
\includegraphics[width=8cm]{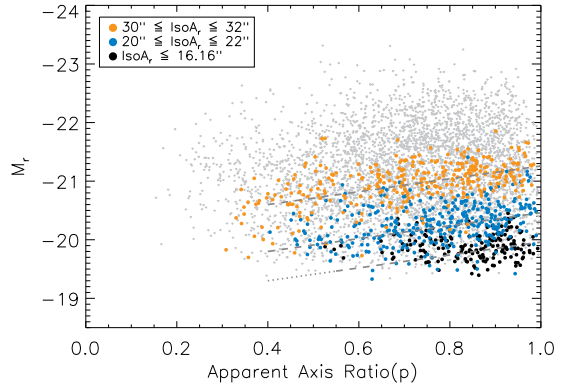}
\caption{The sample bias caused by the optical brightness criterion.
Gray dots in the background indicate all galaxies within
$0.05\leq z \leq 0.06$. Galaxies of different apparent isophotal major-axis
sizes are grouped into 3 bins: orange dots denote the largest galaxies.
Three dashed lines show the trend that rounder galaxies are brighter for a
fixed apparent axis ratio. Faint flat small galaxies
($\mathrm{IsoA_r} \lesssim 16.2''$), near black dotted line,
are missed in this kind of survey (see text).} \label{fig3}
\end{center}
\end{figure}

A complete volume-limited sample is crucial for this study because
it has a direct effect on the intrinsic shapes of galaxies.
But SDSS provides spectroscopic information only for the galaxies of
$r<17.77$ ; hence, our sample cannot be free from luminosity bias.
To investigate this, we need to know how the ARD varies with the size
of major axis.
Fig. \ref{fig3} shows the trend that the luminosity gradually
increases with increasing minor axis for a fixed major axis size.
This effect is clearer for the larger galaxies (red and blue dots)
but less clear for smaller galaxies (black dots) probably because of the
magnitude limit. We also denote the general trends of the three different
luminosity classes with three lines.
It is clear that small faint flat early types are more easily missed,
and so the luminosity limit biases the apparent ARD.
We have decided to exclude galaxies with major axis radius $\mathrm{IsoA_r}$
smaller than 16.2''. This however has a tendency of removing distant
faint galaxies from the sample. In order to alleviate this problem,
we construct a volume-limited sample by selecting close galaxies within
$0.05\leq z \leq 0.06$.
Obviously, no redshift dependence is considered important here.
Note that Odewahn et al. (1997) already reported that there is no
significant difference in the apparent ARD between distant and nearby samples.
Constraining the absolute magnitude range, statistical
Kolmogorov-Smirnov (KS) test confirms that distant
($0.097 \leq z\leq 0.1$) and close ($0.05 \leq z \leq 0.06$) galaxies share
the same parent ARD with a 99\%-level confidence.
Therefore, the results from the analysis on our close
($0.05 \leq z \leq 0.06$) sample likely holds for a larger redshift range.

\subsubsection{fracDev Dependence}

It is also of interest whether our main selection criterion
{\it fracDev} $\geq 0.95$ influences the ARD.
S. Joo (priv. comm.) performed an independent classification of
early types using {\it fracDev} and visual inspection, for two different
{\it fracDev} limits, {\it fracDev} $ \geq 0.95$ and
$0.50 \leq ${\it fracDev} $\leq 0.95$.
He pointed out to us that the higher {\it fracDev} criterion misses some
early types,  mainly flatter galaxies.
For example, if one uses lower {\it Fracdev} limit,
the fraction of relatively flatter (b/a $<$ 0.6) galaxies with $M_r >-20.5$
would be 5.9\%, while the fraction for higher {\it fracDev} limit is 3.9\%.
In this regard, our sample does
not represent the entire early-type galaxy population but is
slightly biased towards rounder galaxies.
We decided, however, not to worry about this,
first because the exact {\it fracDev} criterion for early types is unclear,
and second because we are for the moment more interested in the methodology.

\subsection{Test Sample}

\begin{figure}
\begin{center}
\includegraphics[width=8cm]{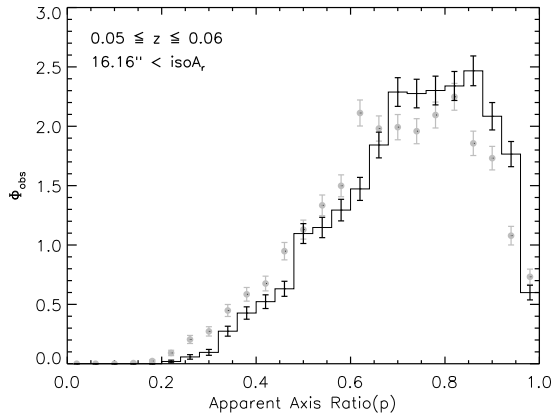}
\caption{Our final SDSS sample compared to the previous APM sample of Loveday (1996)
(Gray dots).
We derive a relatively-complete sample concerning luminosity and major axis radius.
} \label{fig4}. \end{center}
\end{figure}

Our final sample of 3,922 galaxies is chosen with redshift criterion
($0.05 \leq z \leq 0.06$) and major axis criterion
($\mathrm{IsoA_r} > 16.16''$).
We remove a small number of relatively faint outliers of $r > 17.5$
to ensure reasonable image quality.
But this has no impact on our results at all.
We also exclude 1,072 galaxies from the 4,994 galaxy sample because
they appear
to be spiral contaminants or severely-distorted in the visual inspection.
The ARD for final sample is shown in Fig. \ref{fig4}.
For the purpose of comparison, we also plot the apparent ARD from
APMBGS data (Loveday 1996).
We bin the data by the size roughly drawn from the Izenman method
(1991), $aIQRn^{-1/3}$,  based on the total number $n$ and interquartile
range (IQR) where $2.0\leq a\leq2.5$ (Izenman 1991).
The peak around $p=0.8$ is noteworthy.

\section{INTRINSIC AXIS RATIO DISTRIBUTION FOR VOLUME-LIMITED SAMPLE}

It was demonstrated in Fig. \ref{fig2} that the apparent ARD is not uniform
even though the intrinsic distribution is uniform.
To extract the intrinsic shapes of our galaxies,
we use composite models of O, P, T types based on the
Franx et al. (1991) classification scheme (Fig. \ref{fig1}).
In this section, we compare the observation with model distributions using two strategies.

\subsection{Uniform distribution of intrinsic axis ratio}
We investigate the possibility of uniform intrinsic distribution.
To measure the goodness of the fit, we use reduced $\chi^2$,
and this is constructed by considering Poisson error.
Using the volume-limited sample from \S 3,
we try to derive the fractions of the O, P, T types in the combination
that best reproduces the observed ARD.
We can express this with weight($W$) of each type.

\begin{equation}
\Phi_{o}W_{o}+\Phi_{p} W_{p} + \Phi_{t} W_{t} = \Phi_{obs}
\label{eq5}
\end{equation}
\begin{equation}
\textrm{where}~~\textrm{apparent ARD}~~ \Phi_{i}= \sum_{\beta}\sum_{\gamma}\Psi_i(\beta,\gamma)\nonumber
\end{equation}
Index $i$ indicates each type, and $\Psi$ is the apparent ARD
for certain axis ratio $(\beta,\gamma)$.
Weight directly reflects the frequency of each type.
The best solution that makes a minimum $\chi^2$ is $W_o:W_p:W_t=0.09:0.00:0.91$
but for a very poor statistic ($\chi_{red}^2\sim50$).
Hence, {\em no linear combination of OPT types with random
ARD reproduces the observed ARD of our sample!.}

\subsection{Gaussian distribution of intrinsic axis ratio}

Since the random intrinsic ARD fails to reproduce the observed data,
we adopt Gaussian distribution.
Although previous studies, which allow arbitrary distribution, could
estimate the preference of axis ratio, it is nearly impossible to quantify
the fraction of oblate, prolate and triaxial type.
In this respect, Gaussian is the best distribution to test the fraction
and perference of axis ratios.
The Gaussian distribution can be written as
\begin{equation}
F_{gau}(x;\mu, \sigma)=\frac{1}{\sqrt{2\pi}\sigma} \exp{[-\frac{(x-\mu)^2}{2\sigma^2}]}
\end{equation}
where $\mu$, $\sigma$ correspond to the mean position and width of Gaussian distribution.
Then preferences for oblate and prolate galaxies can be expressed as
\begin{eqnarray}
\Phi_{o}= \sum_{\beta}\sum_{\gamma}F_{gau}(\gamma;\mu_o, \sigma_o)\Psi_o(\beta,\gamma) \\
\Phi_{p}= \sum_{\beta}\sum_{\gamma}F_{gau}(\beta;\mu_p, \sigma_p)\Psi_p(\beta,\gamma)\nonumber
\end{eqnarray}

For triaixal galaxies, we use two Gaussian weights following the Lambas et al. (1992) approach.
\begin{equation}
\Phi_{t}= \sum_{\beta}\sum_{\gamma}F_{gau}(\beta;\mu_{t,\beta}, \sigma_{t,\beta})F_{gau}(\gamma;\mu_{t,\gamma}, \sigma_{\gamma})\Psi_t(\beta,\gamma)
\end{equation}
for $\{(\beta,\gamma)~|~T_{i,min} \leq (\frac{1-\beta^2}{1-\gamma^2})< T_{i,max}\}$ where $0.2 \leq \gamma \leq \beta \leq 1.0$.

To reproduce the observed ARD above, we imposed specific intrinsic axis ratios and
sum over three probability distributions after multiplying their weight factors.
We should note that these weight factors are different from Gaussian weight in Eqn. \ref{eq5},
and recall that this provides a simple framework
where we can investigate the fraction of each type.

\begin{figure}
\begin{center}
\includegraphics[width=8cm]{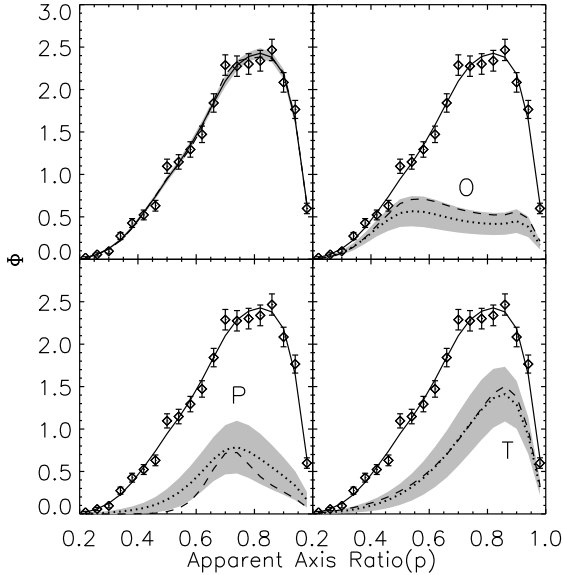}
\caption{Apparent axis ratio distributions based on our Gaussian weight
scheme. The OPT-combined best fit is shown by the dashed line, and the
corresponding component fits (O, P, T) are shown.
We exhibit the average value with the dotted line and 1$\sigma$ deviation of
each model that satisfies $\Delta \chi^2 \leq 1$ as shaded region.
Note that triaxial galaxies are the dominant component.
} \label{fig5}
\end{center}
\end{figure}

\begin{figure}
\begin{center}
\includegraphics[width=8cm]{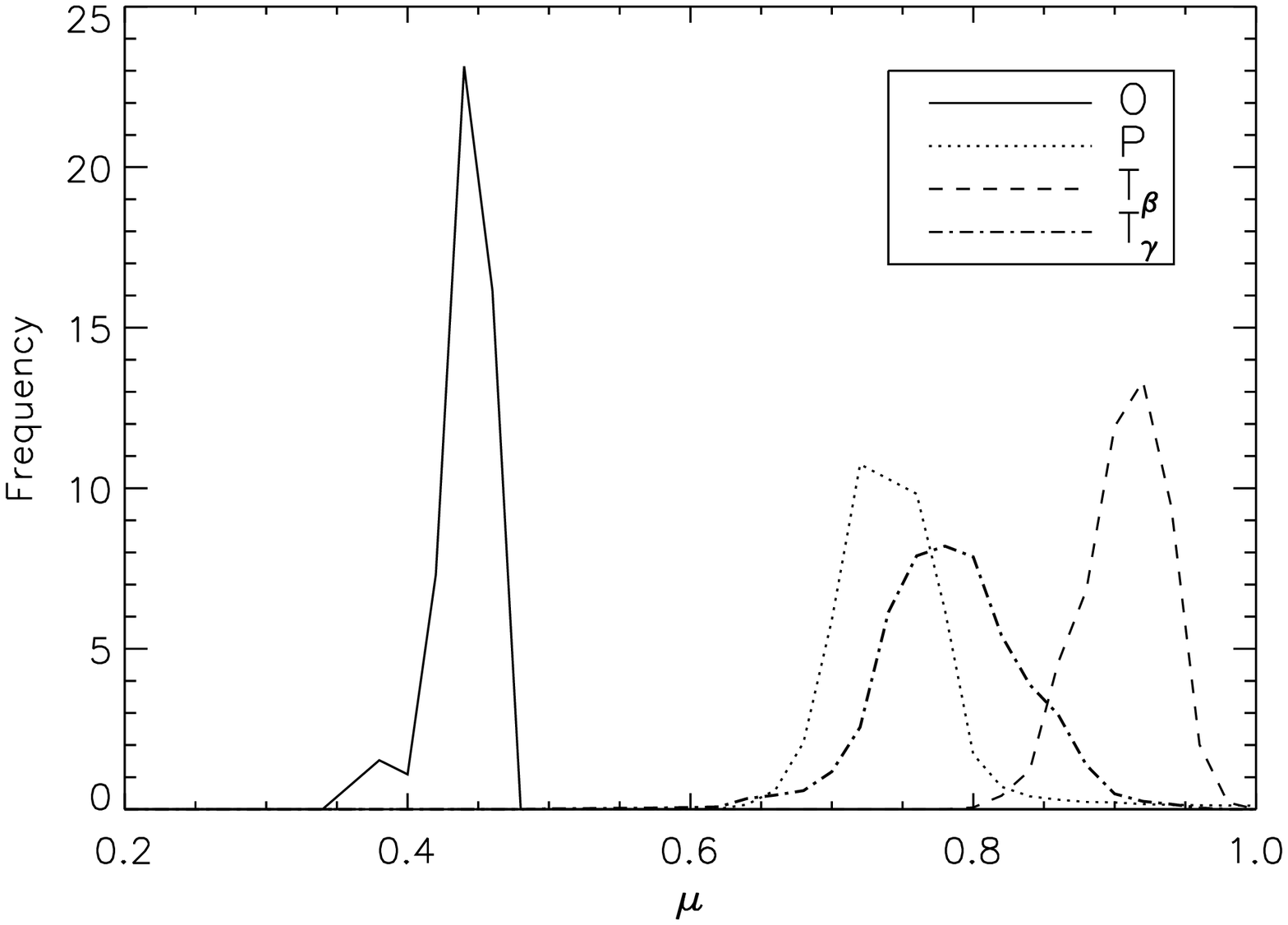}
\includegraphics[width=8cm]{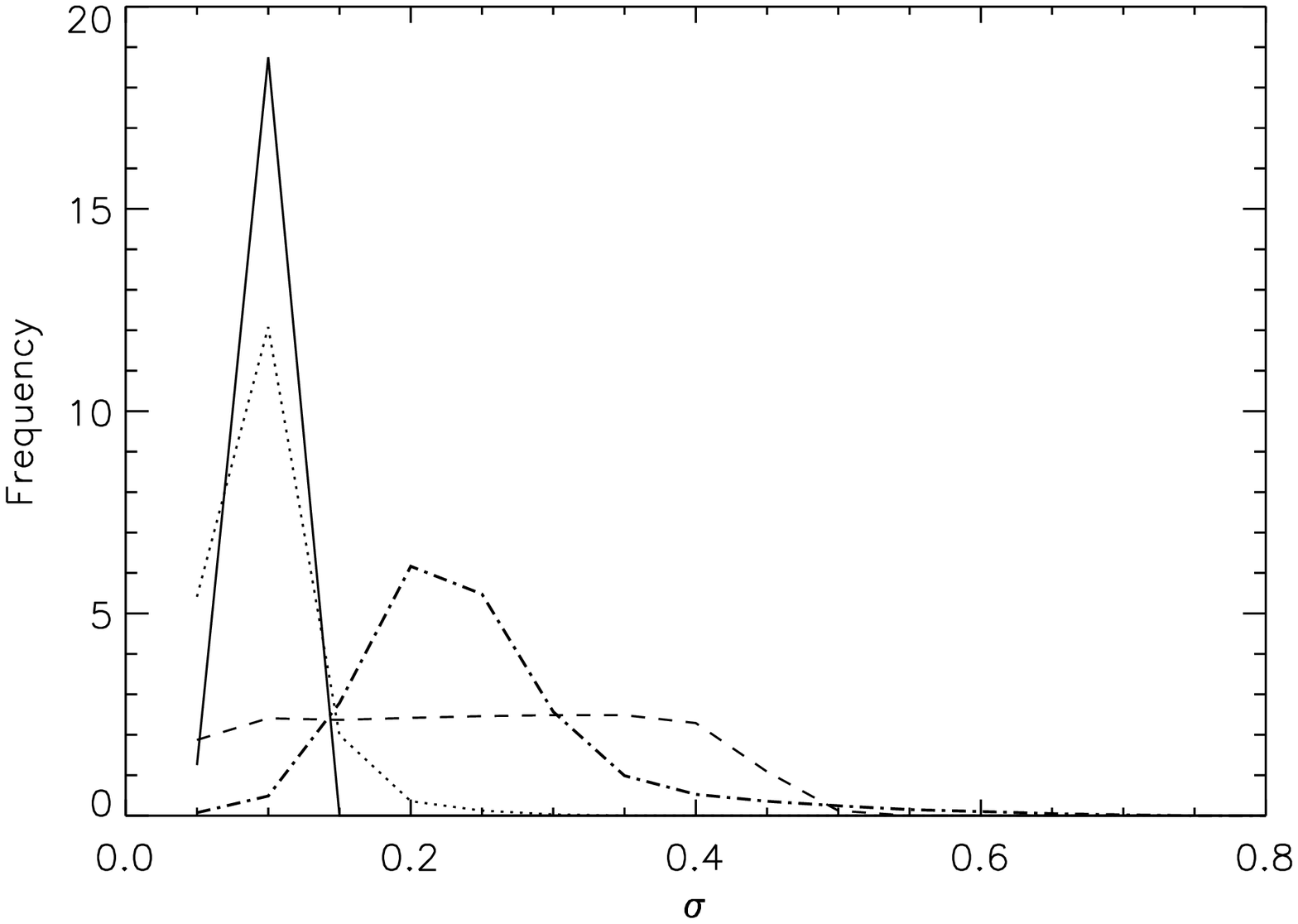}
\caption{{\it Left:} Preferred intrinsic axis ratio within a 1 $\sigma$ range.
We confirm that the preferred axis ratio of each type converges at some axis
 ratio.
{\it Right:} Same as top, but for 1$\sigma$ width of Gaussian weight.
$\sigma_o$, $\sigma_p$ and $\sigma_{t,\gamma}$ shows good convergence
while $\sigma_{t,\beta}$ is widely accepted.
} \label{fig6}
\end{center}
\end{figure}

With this approach we find good matches.
The best-fit model that yields minimum $\chi^2 \approx 1$ is
$(\mu_o,~\mu_p,~\mu_{t,\beta},~\mu_{t,\gamma})=(0.46,~0.72,~0.92,~0.74)$, and
$(\sigma_o,~\sigma_p,~\sigma_{t,\beta},~\sigma_{t,\gamma})=(0.1,~0.05,~0.1,~0.3)$.
For this case, we found that the fraction of each type is
$\textrm{O:P:T}=0.35:0.18:0.47$.
However, our parameter space is so complicated that
the minimum $\chi^2$ model may not represent the most meaningful result.
Instead, the statistical properties of all possible models with reduced
$\chi^2$ values within 1$\sigma$ range
( $\Delta \chi^2 = \chi^2_\nu - \chi^2_{min} \leq 1$) are
more meaningful because they all show good agreements with the observation
(Fig. \ref{fig5}).
Our results show that the triaxial component is dominant around the
high axis ratio, while oblate also plays an important role in the low axis
ratio region.  Statistically, the total fraction of each type is
$\textrm{O:P:T}=0.29^{\pm0.09}:0.26^{\pm0.11}:0.45^{\pm0.13}$ in 1 $\sigma$ range,
which suggests that triaxial early-types are most common.
For comparison to the ``best-fit model'', the $\chi^2$ space of the 
``good models'' ($\Delta \chi^2 <1$) show a convergence at 
a slightly different configuration 
($\mu_o$, $\mu_p$, $\mu_{t,\beta}$, $\mu_{t,\gamma}$) 
= (0.44, 0.72, 0.92, 0.78) as shown in Fig. \ref{fig6}.
We believe that this is a more statistically-representative result.

In the similar simulation of Lambas et al. (1992),
the optimal solution for fitting the apparent ARD from APMBGS data
with two-dimensional Gaussian had $\mu_{t,\beta}=0.95$, $\mu_{t,\gamma}=0.55$,
$\sigma_{t,\beta}=0.35$, $\sigma_{t,\gamma}=0.2$.
Likely reasons for the difference between their results and ours are:
(1) they considered only triaxial and used different classification scheme;
and (2) the observed ARDs are slightly
different. They used smaller values of $\mu_{t,\gamma}$ (i.e., flatter)
than ours probably in order to fit the low axis ratio
regions without considering oblate and prolate elements.
On the other hand, Ryden's (1992) results that produce the best-fit model
with $\mu_{t,\beta}=0.98$, $\mu_{t,\gamma}=0.69$ and $\sigma_{t}=0.11$ are
closer to our results; but, since their observed ARD was suppressed in the
low axis ratio region, their best-fit triaxials were rounder than those
in the Lambas et al. fit.

\section{INTRINSIC ARD FOR DIFFERENT LUMINOSITIES}

\begin{figure}
\begin{center}
\includegraphics[width=8cm]{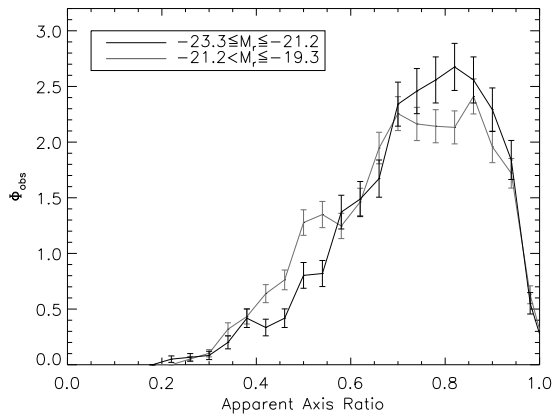}
\caption{Apparent ARDs for the two different luminosity samples.
The solid line and dashed line correspond to ``luminous'', and
``less luminous'' samples, respectively.
The ``luminous'' sample is rounder than ``the less luminous'' sample
in general. }
 \label{fig7}
\end{center}
\end{figure}

We investigate whether two luminosity classes mentioned in \S 1
(Bender 1988, Kormendy \& Bender 1996) have different intrinsic shapes
by comparing their best-fit (within 1 $\sigma$) model OPT ratios.
First, we divide galaxies into ``luminous'' ($M_r \leq -21.2$) and
``less luminous'' ($M_r > -21.2)$) groups.
Rest \& van den Bosch (2001) found a division at $M_B\sim-20$,
which corresponds to $M_r\sim -21.2$ using the transformation of Smith et al.
(2002) and the typical color for early-types ($B-V\sim0.9$).
As seen in Fig. \ref{fig7}, the ``luminous'' sample exhibits a larger
number of round galaxies.
We derive the weight and preference of intrinsic axis ratio for each type.

\begin{figure}
\begin{center}
\includegraphics[width=8cm]{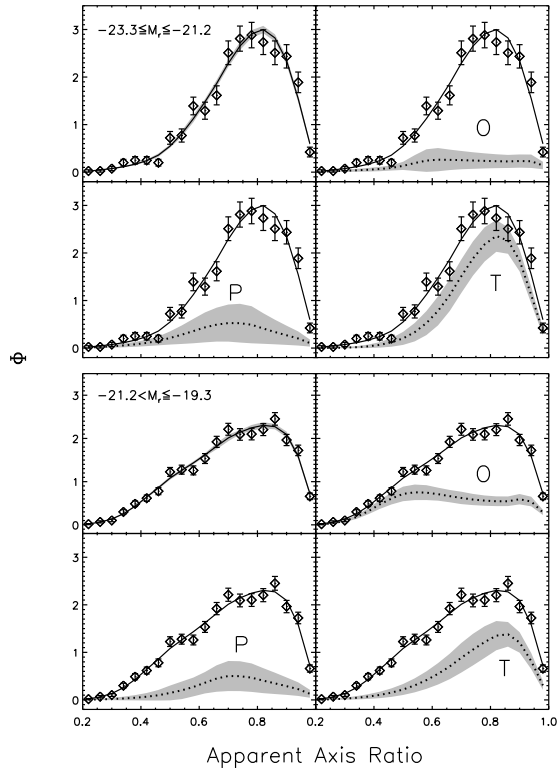}
\caption{Same as Fig. \ref{fig5} but for two different luminosity groups (see text).
Upper panels show the ARDs for the ``luminous'' sample and lower panels are for the
``less luminous'' samples. The majority of the ``luminous'' sample are likely triaxial,
while a large amount of oblate types still exist in the ``less luminous'' sample.}
\label{fig8}
\end{center}
\end{figure}

We first focus on the weight of each type between the two groups.
Averaged over the region of the 1$\sigma$ range ($\Delta \chi^2 \leq 1$),
the OPT weights for the ``luminous'' and ``less luminous'' samples are as
follows.
\begin{eqnarray}
\textrm{O:P:T}&=&0.13^{\pm0.08} : 0.20^{\pm0.13} : 0.67^{\pm 0.13} {\rm (luminous)}\nonumber\\
\textrm{O:P:T}&=&0.38^{\pm 0.08} : 0.18^{\pm0.10} : 0.43^{\pm0.11} {\rm (less~luminous)}\nonumber
\end{eqnarray}
This implies that luminous early-types are likely triaxial,
while there still exists a large amount of oblate galaxies in the
``less luminous'' sample. We display in Fig. \ref{fig8}
how each type can be viewed in the sky.
Note that the galaxies with a high apparent axis ratio are likely
triaxial regardless of their luminosity, and oblate galaxies are
more common in the ``less luminous'' sample. This may indicate that
different formation process between the two samples. We will discuss this
in greater detail in \S 8. The dichotomy, if real, might be explained
by the argument presented by Valluri \& Merritt (1998) involving
the central supermassive black hole and the crossing time difference between
the bright and faint ellipticals.

However, regarding preferences of intrinsic axis ratio ($\mu$) and
Gaussian widths ($\sigma$), there is no clear disparity between the two
samples.
The ``luminous'' sample shows a preference for $\mu_{t,\beta}=0.90$, 
$\mu_{t,\gamma}=0.70$, $\mu_p = 0.75$, while $\mu_o$ cannot be constrained
because of the minimal contribution from oblates. 
Similarly, for the ``less luminous'' sample, $\mu_o=0.45$, $\mu_p = 0.70$, 
$\mu_{t,\beta}=0.90$ and $\mu_{t,\gamma}=0.70$ -- 0.75 are derived.

McMillan et al. (2007) pointed out that the axis ratios of equal-mass
merger remnants can be different for differing merging conditions.
Then, our result (no significant difference between the two samples
in terms of the axis ratios preferred) could be interpreted as
a lack of significant difference in the merger history between the samples.
However, if it takes numerous merging events to build elliptical galaxies,
the memory of the past merger history could easily be buried.
The last major/minor merger event on the other hand could still be important.

\section{DEPENDENCE ON ENVIRONMENT}

Within the context of $\Lambda$CDM scenario, galaxies build up
hierarchically via galaxy mergers and interactions.
In this regard, the density-morphology relation (Dressler 1980)
may reveal the significance of environment for galaxy formation and
evolution.
We investigate whether the intrinsic shapes of early-types are also
related to environment. We used Joo Heon Yoon's density parameter
($\rho$) for the local density of early-type galaxies
(priv. comm.).
Yoon's density parameter ($\rho$) is a measurement of crowdedness counting all
the neighboring galaxies with a Gaussian-weighting scheme in proportion to the
distance between the galaxies. Yoon's measurements are improved over 
Schawinski et al.'s
primarily by including more candidate member galaxies assuming that early-type
galaxies obey optical color-magnitude relations.
Surprisingly, 
we did not find any notable impact of the density parameter on the
apparent ARD. Even between the two extreme sub-samples
(representing fields and dense clusters), the apparent ARDs are found to 
share the same parent distribution via a K-S test.
It should be noted that Ryden (1993) reported a similar result, finding
no significant difference in the ARD in her sample of brightest cluster
galaxies (BCGs).

It is interesting to note that semi-analytic models also show
consistent results.
Khochfar \& Silk (2006) assert that the stellar
properties of merging remnants of massive galaxies above the
characteristic mass ($\sim 3\times10^{10} \mathrm{M}_{\odot}$;
Kauffmann et al. 2003) are not much different between the field and
cluster environments.
Considering the brightness of our sample galaxies $M_r = -19$ through $-23$,
and assuming a stellar mass-to-light ratio 5,
most of our galaxies would exceed the characteristic mass.

\section{Limitations}

In this section we investigate the limitations of our approach
by comparing our results with different classification schemes.
A robust result that does not strongly depend on the choice of schemes
would ensure the usefulness of the method.
Table 1 and Fig. \ref{fig10} show the three cases defined as
M1, M2, and M3, where M2 is the Franx et al. classification scheme we used
in this study.

\begin{figure}
\begin{center}
\includegraphics[width=8cm]{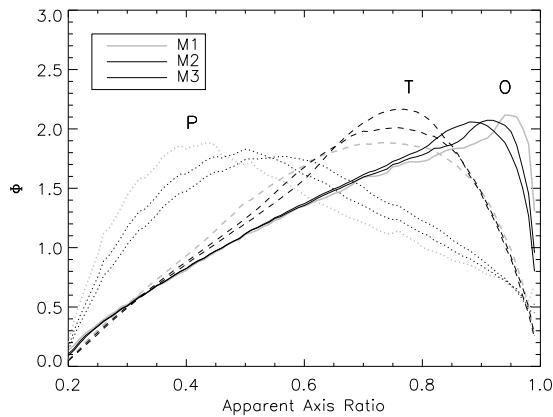}
\caption{Same as Fig. \ref{fig2} but for different classification schemes.
The oblate and prolate types are shifted towards the triaxial peak as
the classification scheme becomes more generous for oblate and prolate.}
 \label{fig9}
\end{center}
\end{figure}

\begin{figure}
\begin{center}
\includegraphics[width=8cm]{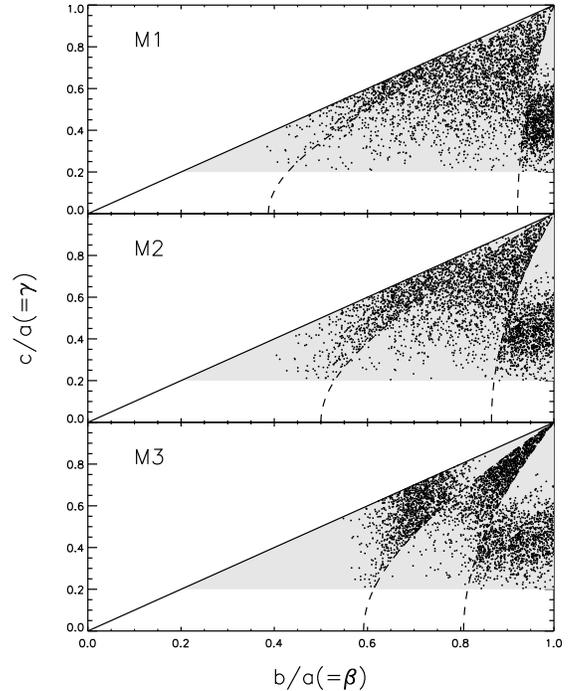}
\caption{A mock intrinsic distribution for a sample of 4000 galaxies
based on the values in Table 2. This figure shows how intrinsic ARD
changes with classification scheme (see text).}
 \label{fig10}
\end{center}
\end{figure}

In Fig. \ref{fig9} we show the apparent ARDs for the three cases
assuming a uniform intrinsic ARD.
The peak of each type can change slightly with the classification scheme,
but the overall behavior is still retained.
In addition, the goodness of fit assuming uniformly-distributed axis ratios
is still poor for other cases ($\chi^2_{red} \sim 50$; see \S 4.1).

To further constrain the Gaussian parameters, we tested the volume-limited
sample using three different classification schemes.
If the dispersion of the preferred axis ratio is too large,
no fitting distribution can maintain its full information of the
assumed distribution.
For example, a Gaussian distribution of oblate galaxies parameterized
with $\mu$ and $\sigma$ (in the $\gamma$ direction)
would inevitably have galaxies lying outside the
oblate territory by the classification scheme (Figure 1): mostly into
triaxial. This is particularly so
when $\sigma$ is large and for triaxial ellipticals due to the type definition.
Nevertheless, our exercise showed that the preferred axis ratios and
type weights are robustly derived by assuming Gaussian distribution.

In Table 2, we present the preferred axis ratios ($\mu$) corresponding to
the most probable case showing reliable convergence and the Gaussian width
($\sigma$) for three different classification schemes.
The results on $\mu$ and $\sigma$ appear to be reasonably consistent 
between M1 and M2, but M3 shows a larger difference.
The slight difference in the OPT fractions derived is easy to understand.
For example, the fractions of oblate and prolate in the M3 case are
greater than those in M1; this is natural because there are more
possible oblate or prolate configurations in M3 than in M1.
In Fig. \ref{fig10}, we show a mock intrinsic ARD for a sample of
4000 galaxies simulated with the values in Table 2.
We first populate the $4000 \times W_{\rm type}$ galaxies for each type using
Gaussian distribution. Note that oblate and prolate populations have 
Gaussian distribution along only one (shortest) axis;
oblates (prolates) have Gaussianity on $\gamma$ ($\beta$).
Fig. \ref{fig10} demonstrates how the intrinsic shapes of galaxies derived
change with classification scheme. 
Once again, M1 and M2 are similar, but not M3.
This demonstrates the currently unsatisfying situation that
the results of our OPT analysis can be sensitive to the choice of the
OPT classification scheme.
In this sense, it is critical to use a classification scheme that
is more physically motivated from kinematic requirements.
If numeric divisions on axis ratios are still the easiest scheme,
the OPT classification could be determined from models.
For example, an ensemble of stellar orbits dominated (by a certain value)
by oblate stellar orbits may have a characteristic range in $c/b/a$ axis
ratios which can then serve as the unique criterion for ``oblate''.

\begin{table}
\begin{center}
 \caption{ Three different classification schemes to test the robustness of our OPT analysis}
\begin{tabular}{ c c c c}\hline \hline
Model & Oblate & Prolate & Triaxial \\
\hline
M1 & 0 $\leq$ T $<$ 0.15 & 0.85 $\leq$ T $\leq$ 1.0 & 0.15 $\leq$ T $<$ 0.85 \\
M2 & 0 $\leq$ T $<$ 0.25 & 0.75 $\leq$ T $\leq$ 1.0 & 0.25 $\leq$ T $<$ 0.75 \\
M3 & 0 $\leq$ T $<$ 0.35 & 0.65 $\leq$ T $\leq$ 1.0 & 0.35 $\leq$ T $<$ 0.65 \\
\hline
 \end{tabular}
 \end{center}
\end{table}

\begin{table}
\begin{center}
 \caption{The preferred axis ratios and fractions of OPT types derived}
\begin{tabular}{ c c c c c c c c c c}\hline \hline
Model & $\mu_o$  & $\sigma_o$ & $\mu_p$ & $\sigma_p$ & $\mu_{t,\beta}$ & $\sigma_{t,\beta}$ & $\mu_{t,\gamma}$ & $\sigma_{t,\gamma}$\\
\hline
M1 & 0.44 & 0.1 & 0.70 & 0.1  & 0.92 & (0.3)\footnote{poorly constrained} & 0.74&0.2 \\
M2 & 0.44 & 0.1 & 0.72 & 0.1  & 0.92 & (0.25)\footnote{poorly constrained}& 0.78&0.2 \\
M3 & 0.44 & 0.1 & 0.72 & 0.05 & 0.96 & 0.1                                & 0.86&0.25\\
\hline
 \end{tabular}

\begin{tabular}{c c c c}\hline\hline
Model & $ W_o $ & $ W_p $ & $ W_t $\\
\hline
M1 & 0.24 $^{\pm 0.10}$ & 0.15 $^{\pm 0.09}$ & 0.61 $^{\pm 0.11}$ \\
M2 & 0.29 $^{\pm 0.09}$ & 0.26 $^{\pm 0.11}$ & 0.45 $^{\pm 0.13}$ \\
M3 & 0.32 $^{\pm 0.07}$ & 0.31 $^{\pm 0.11}$ & 0.37 $^{\pm 0.12}$ \\
\hline
 \end{tabular}
 \end{center}
\end{table}

\section{CONCLUSIONS AND DISCUSSION}

We have investigated on the intrinsic shape distribution of early-type
galaxies. We show that the de-projection results can be
affected by the details of the selection criteria of the observed data.
We use isophotal major radius, luminosity, redshift and $fracDev$ combined
in order to minimize sample biases.
We have constructed a volume-limited sample of 4,994 early-type galaxies.
In addition, since even the high $fracDev$ criterion ($>0.95$) does not
completely remove the contamination of non-early-type galaxies,
we performed visual inspection on the galaxies and finally selected 3,922.

We use the projection probabilities analytically calculated to
determine the ratios of oblate, prolate and triaxial (OPT) types.
We found no combination of randomly-distributed OPT types
matches the observed ARD!

We have tested a hypothesis of Gaussian distribution in axis ratios.
The results show excellent fits to the observed ARD from our volume-limited
sample.
In particular, our results show that triaxial is the most common type
($W_t=45\%\pm12$), and they are round in general
($\mu_{t,\beta}=0.92,~ \mu_{t,\gamma}=0.78$).
On the other hand, oblate prefers a more flattened axis ratio
($\mu_o=0.44$) and accounts for approximately 29\% of all early-types.
The prolate show a peak at $\mu_p=0.72$ with a comparable fraction to
that of oblate (26\%).

Recent numerical N-body simulations suggest that the shapes of
dark matter halos favor triaxial or prolate over oblate
(Dubinski \& Carlberg 1991, Barnes 1992, Jing \& Suto 2002,
Bailin \& Steinmetz 2005, Novak et al. 2006).
This is consistent with our result at least in the sense
that triaxial is favored, while a small discrepancy is found in the
values of the preferred intrinsic axis ratios.
Dubinski \& Carlberg (1991) proposed that
dark matter halos are triaxial with $<\beta> = 0.71$
and $<\gamma> = 0.50$, respectively, which is somewhat more flattened
than we derived.
In order to reproduce the observed ARD using these preferences for triaxial
type, specifically-designed distributions for oblate and prolate types
would be needed. This is obviously contrived.
Interestingly, our results are more comparable with the simulation results of
Bailin \& Steinmetz (2005) that suggested
$\gamma=0.6\pm0.1$ and $\beta=0.75\pm0.15$ for dark matter haloes.
It should however be noted that the mass distribution of halos with
cooled baryon can be different from those of pure dark matter halos
(Gnedin et al. 2004). Novak et al. (2006) also pointed out the
disparity in the halo shape between the stellar and dark matter components.
If this is true, the apparent agreement between the axis ratios derived
in our study and those from the dark matter simulations may not be
significant. Further investigations are called for.

In the context of merging, merger remnants are mainly affected not only by
initial orientation but also by the mass ratio of the
merging galaxies (Naab \& Burkert 2003; Khochfar \& Burkert 2006).
Bright early-type galaxies are often supposed to be formed via numerous
merging events in the $\Lambda$CDM cosmology.
Hence, we should investigate all possible parameter space in order to make
a realistic comparison with observation.
In addition, because elliptical-elliptical mergers, as well as spiral-spiral
mergers, can lead to form an elliptical galaxy, the parameter space is large.
Furthermore, there appear to be two distinct classes of early-type
galaxies separated by their luminosity (Bender 1988, Kormendy \& Bender 1996).
These studies indicate that luminous galaxies are typically supported by
anisotropic velocity and have boxy isophotes while faint early-types
show disky isophotes. On this basis, we search for a clue on their
formation history using the OPT parameters we derived.
Our results indicate that the
``luminous'' sample ($M_r < -21.2$) are mostly triaxial ($W_t=67\%\pm13$)
while the ``less luminous'' sample have a large number of oblate types
($38\%\pm8$). Interestingly, no clear difference in the
preferred axis ratios of OPT types between the two samples.

Khochfar \& Burkert (2003; 2005) suggest that the origin of the two classes
could come from the different types of progenitor being merged (c.f., Valluri
\& Merritt, 1998). Their semi-analytic models suggest that
bright early types tend to form from  elliptical-elliptical mergers.
Naab et al. (2006) also pointed out that spiral-spiral mergers cannot
reproduce the observed fraction of anisotropic early-types, and that
boxy, anisotropic system can form by binary mergers of early-type galaxies.
If early-type mergers are the main channels to form bright anisotropic
galaxies, and if our triaxial galaxies correspond to anisotropic
early-types while oblate types are rotationally-supported systems,
our results show reasonable agreement with Naab et al. (2006) in terms of
the ratio of anisotropic to the total number of early-type galaxies.
The fraction of the anisotropic galaxies in the ``luminous'' and the
``less luminous'' samples are around 0.8 and 0.4, respectively (Naab et al.
2006), which are arguably similar to our results $W_t=0.67\pm0.13$ and
$W_t=0.43\pm0.11$.

Since different merging events are proposed to 
result in different configurations for
early-type galaxies, it is also of interest to investigate
whether there is connection between environment and galaxy shape.
We do not find a clear dependence on the local galaxy density.
Considering that the galaxy number density parameter generally represents
the dark matter halo potential for the galaxy cluster,
this may imply that such a cluster-scale environment might have little
effect to the intrinsic shapes of individual galaxies.
This is consistent with the results from the semi-analytic study of
Khochfar \& Silk (2006).

Our analysis has caveats. The quantitative aspects of our simulation
results depend on the OPT classification scheme. Besides,
when the fraction of a type (oblate, prolate, or triaxial) is small,
our method fails to constrain the preferred values of mean axis ratios and
Gaussian widths. With the improvement on these caveats and detailed study
of N-body simulations, we hope our work can provide a simple framework
to investigate the formation history of early-type galaxies.

\acknowledgements

We are very grateful to Seok-Joo Joo who kindly provided advice on the
galaxy morphology, and Chang H. Ree for many useful comments on the
data selection.
This work could not have been the same without their help.
We thank Sadegh Khochfar, Tim de Zeeuw and James Binney for constructive comments
in the early stage. We are greatly indebted to the anonymous referee for
numerous constructive criticisms and clarification.
This work was supported by grant No. R01-2006-000-10716-0 from the Basic
Research Program of the Korea Science \& Engineering Foundation.

\end{document}